\RequirePackage{lineno} 

\documentclass{article}
\usepackage{fancyhdr}

\usepackage[round]{natbib}
\usepackage{graphicx}
\usepackage{multirow}
\usepackage{color}
\usepackage{amssymb}
\usepackage{amsmath}
\usepackage{enumerate}
\usepackage{lineno}
\usepackage{caption}
\usepackage{textcomp}
\usepackage{gensymb}
\usepackage{upgreek}
\usepackage{float}
\usepackage{lipsum}
\usepackage{epstopdf, epsfig}
\usepackage{authblk}
\usepackage{rotating}

\usepackage[letterpaper, margin=1.0 in]{geometry}

\begin{document}

\title{Estimating sea ice properties from wave observations in sea ice}
\author[1]{J.J. Voermans}
\author[2]{J. Rabault}
\author[3]{A. Marchenko}
\author[4]{T. Nose}
\author[4]{T. Waseda}
\author[1]{A.V. Babanin}

\affil[1]{Department of Infrastructure Engineering,	University of Melbourne, Melbourne, Australia}
\affil[2]{Norwegian Meteorological Institute, Oslo, Norway}
\affil[2]{Arctic Technology Department,	The University Centre in Svalbard, Longyearbyen, Norway}
\affil[4]{Graduate School of Frontier Sciences,	The University of Tokyo, Kashiwa, Chiba, Japan}
\setcounter{Maxaffil}{0}
\renewcommand\Affilfont{\itshape\small}
\date{}
\maketitle

\begin{abstract}
	The Marginal Ice Zone is a highly dynamic region where the atmosphere, ocean, waves and sea ice meet. Waves play a fundamental role in this coupled system, but our progress in understanding wave-ice interactions is currently hindered by the lack of observations of sea ice properties in-situ. In this study we aim to estimate the ice thickness and effective elastic modulus of sea ice passively using observations of waves in ice obtained from vibrations sensors. Specifically, we use three low-cost geophones deployed in triangular formation with sides of about 200 m on fast ice. The focus here is on three major wave events that were recorded, each consisting of initial high frequency vibrations with a frequency at around 10 Hz, followed by low frequency dispersive waves within a frequency range of 0.08--0.28 Hz. Based on the phase speed of the initial high frequency vibrations, we estimate the purely elastic effective modulus to be 4--4.5 GPa.	By comparing the arrival times of the dispersive low frequency wave events at the geophones to the dispersion relationship of waves in a thin elastic ice sheet we obtain estimates of the effective elastic modulus in the range of 0.4--0.7 GPa. This is close to the measured effective elastic modulus through cantilever experiments of 0.5 GPa but considerably smaller than the default value of 5.5 GPa currently in use in contemporary wave models. We could not, however, obtain explicit estimates of the ice thickness and effective modulus individually from the measured low frequency dispersive waves as their impacts on the shape of the dispersion relationship are similar in this frequency range. That is, information of one is required to estimate the other. Distinction is only possible for dispersive wave events larger than about 1 Hz, such as for vibrations generated by (thermal) cracking, which are frequently observed in our dataset. Using source triangulation, we find that the major wave events observed in this study originate from the nearby glacier. We suspect that the high frequency vibrations are caused by the calving of the glacier. With our results, we show that low-cost geophones can be used to estimate sea ice properties in fast ice and substantiate that this to be possible on very large ice floes as well. Progress in retrieving sea ice properties from wave measurements is expected to increase the dataset of in-situ sea ice properties observations in number and geographic coverage and improve our understanding of wave-ice interactions.
\end{abstract}

\section{Introduction}
Waves can propagate hundreds of kilometers intro the sea ice before the majority of their energy is attenuated by wave scattering \citep{Montiel16,Kohout08} and wave dissipation processes, including turbulence \citep{Kohout11,Voermans19,Herman21}, internal friction \citep{Wang10}, overwash \citep{Nelli20} and ice-floe collisions \citep{Herman19b,Rabault19JFM,Loken22}. The rate of attenuation is largely determined by the properties of the ice where, in general, thicker and larger ice floes result in larger attenuation of wave energy. When the wave steepness is sufficiently high, waves can break the ice \citep{Dumont11,Voermans20}, thereby reducing its attenuation capacity and allowing waves to penetrate further into the ice cover, ultimately leading to increased break-up activity and an increase of the Marginal Ice Zone (MIZ) width \citep{Asplin12,Collins15}. Wave steepness can be further impacted by the change in wave group velocity when entering the ice which, in the case of short waves, tends to decrease the wave height and thus wave steepness \citep{Collins17}. Implementation of these complex feedbacks between the waves and the ice in our forecasting models is thus strongly dependent on our collective knowledge of wave attenuation, wave dispersion and wave-induced sea ice break-up in terms of the sea ice properties and, importantly, the availability thereof \citep[e.g.,][]{Kousal22,Boutin20}.

Many theories and models currently exist, with varying degree of complexity and representation of the underlying wave-ice physical processes \citep[e.g., see][for an overview]{Shen19,Squire20,Rogers21}. All these models rely in one way or another on the properties of the ice. Notable sea properties are the ice thickness and elastic modulus, but others include the flexural strength, roughness (both above and below the waterline) and horizontal dimensions of the ice floes. Validation of the theories and models against field experimental observations is however very limited, largely due to the difficulty of obtaining observations of waves and sea ice in the field.

Measuring waves and sea ice properties in the MIZ is hindered by the extreme complexity of this harsh environment which poses logistical and technological challenges. Despite this, significant progress has been made in estimating wave properties in the MIZ due to developments and advancements in low-cost instrumentation \citep[e.g.,][]{Rabault22} and satellite derived products \citep{Ardhuin15,Horvat20}, which already has led to a substantial increase of the availability of in-situ wave observations \citep{Rabault22data}. In contrast, access to observations of the physical and mechanical properties of sea ice remain restricted, with the exception perhaps of sea ice thickness which may be estimated from satellite derived products \citep[e.g.,][]{Pactilea19}, although at relatively coarse spatial resolution and primarily for thin ice. Importantly, the elastic modulus remains a sparingly documented property of sea ice in relation to wave-ice interaction studies.

The elastic modulus $E$ is equal to the ratio of the stress and strain in the ice sheet during elastic behavior \citep{Timco10}. When considering most engineering applications, however, the ice does not behave as purely elastic and delayed elasticity may become important during small loading rates. We may define the total recoverable strain as the sum of the purely elastic strain and the delayed strain, denoted by an effective elastic modulus $E^*$ \citep{Williams13}. \citet{Timco10} suggest typical values of the effective elastic modulus between 1--5 GPa, where \citet{Williams13} opt for a somewhat higher value of $4-7$ GPa, and \cite{Karulina19} slightly lower with observations between 0.3--4 GPa. Due to the absence of observational data of $E^*$, the default value in contemporary wave models is currently 5.5 GPa \citep{WW3DG19} which, by lack of better alternatives, is commonly adopted in wave related studies in the MIZ \citep[e.g.,][]{Boutin18,Williams17,Li21}. This poses two problems. Firstly, it provides an additional tuning parameter to fit model output to the observations, thereby potentially masking the mechanistic output of the models or, in other words, it obscures our qualitative understanding of wave-ice interactions \citep[e.g.,][]{Voermans21b}. Secondly, an unknown elastic modulus leads to quantitative uncertainty in the model output including critical dynamical variables, such as the MIZ width. For example, choosing wave scattering as an attenuation model, \citet{Williams17} showed in their numerical experiments that a doubling of the effective elastic modulus may double the MIZ width during swell events. We note that an uncertainty of $E^*$ by a factor of two (or even more) is far from uncommon as in most wave-ice studies $E^*$ is not measured nor inferred. The sensitivity of the wave-ice interaction models to the properties of the ice, and thus their collective outcome and feedback into the dynamics of the MIZ as a whole, highlights the need for more frequent and accurate observations of sea ice properties.

Traditional methods of measuring sea ice properties tend to be laborious, whether it being sea ice coring to obtain sea ice samples and measuring sea ice thickness, or in-situ cantilever experiments to retrieve effective mechanical properties of the ice \citep{Marchenko20b,Karulina19,Timco10}. Ideally, sea ice properties are inferred from long-term in-situ sensor deployments without the requirement of continuous human presence, which would then allow for higher temporal and/or spatial resolution of such observations. Examples include floe-size distribution from stereo-cameras \citep[e.g.,][]{Alberello19} and ice thickness from ice-mass-balance buoys \citep[e.g.,][]{Richter06}. Measuring the mechanical properties  of ice without physical activity seems less straightforward. The elastic modulus of sea ice, for instance, can be estimated based on observations of the propagation speed of waves in the ice \citep{Stein98,Yang94,Moreau20}. This can be inferred from the phase speeds of flexural waves, compressive and/or shear waves, but typically involves the deployment of a series of seismometers or geophones to track the waves in the spatial domain \citep{Moreau20,Yang94,Oliver54}. This can be an intensive and expensive exercise with some studies using ten to hundreds of vibration sensors to measure the phase speed of waves in ice created by hammer blows, explosions or jumping people. However, the use of active sources is, to some extent, a restriction of the method's generalization as it still requires human presence to generate these active sources. Examples of studies using passive sources do exist, such as \citet{Voermans21b,Sutherland16,Marsan12}, who used accelerometers and seismometers to estimate the ice thickness and/or the elastic modulus from the phase speed of gravity waves, but those will require the presence of gravity waves in the first place and may fail once the ice is broken due to the restrictions in the horizontal length scales of the ice relative to the wave length. In this study, we aim to retrieve sea ice properties from passive wave observations. Specifically, we focus here on the retrieval of sea ice thickness and effective elastic modulus of sea ice from observations of waves in landfast sea ice using three geophones.

\section{Methods}
\subsection{Experimental Setup}
\begin{figure}[ht]
	\centering
	\includegraphics{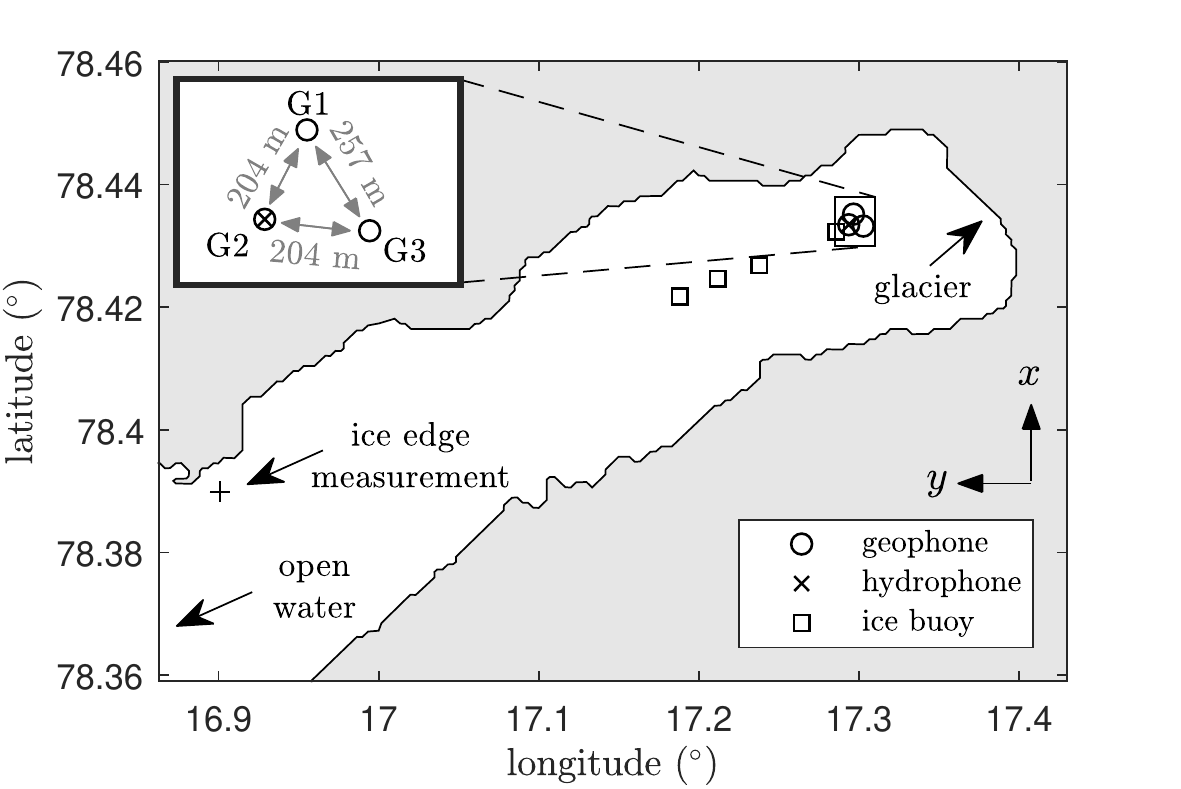}% Images in 100% size
	\caption{Experimental setup in Tempelfjorden, Svalbard. Instrument deployment consists of three geophone logger (also see inset), a hydrophone and four wave-ice buoys. In-situ cantilever experiments were performed near the ice edge, marked by the `+' sign.}
	\label{fig:fig1_experiment_setup}
\end{figure}
To measure waves in sea ice, various instruments were deployed on landfast sea ice from 11--28 February 2022 in Tempelfjorden, Svalbard. The experimental setup consist of multiple ice buoys \citep{Rabault20}, one hydrophone (Audiomoth Dev v1.0.0. with an Aquarian Audio h2a hydrophone) and three geophone loggers. The ice buoys focus on the frequency range between 0.05--1 Hz (i.e., ocean waves), whereas the geophones target sea ice vibrations with frequencies up to 240 Hz. The hydrophone logger was deployed to record the acoustic radiation of ice events underwater, such as cracking and break-up, with a sampling rate of 48kHz. The ice motion loggers were deployed along a line parallel to the main axis of the fjord and the geophones were deployed closest to the glacier-side of the fjord in triangular formation with sides of about 200 m. The distance between the geophones needs to be large enough to record the fastest waves, such as compressive and/or shear waves which have phase speed of a few thousand meters per second, but small enough to limit significant wave transformation and wave attenuation effects. The hydrophone was deployed next to the geophone closest to the open water (see Fig. \ref{fig:fig1_experiment_setup}). In this study no data from the ice buoys were used as the measured wave energy during the measurement campaign was very small.

Ice thickness was measured at the sites of the ice buoys at the time of deployment. The thickness was 35, 40, 47 and 52 cm thick in the direction from the open water towards the glacier, respectively. The thickness did not change noticeable over the duration of the experiment. Air temperatures measured at Svalbard Lufthavn (45 km from the deployment site) varied between -27$^\circ$ and -5$^\circ$C. The local water depth is roughly 40 m at the site \citep{Marchenko13}.

\subsection{Geophone logger}
A custom geophone logger was designed for this project to record sea ice vibrations. While various low-cost and open-source designs are available to record the analog signals of the geophones, such as the Geophonino of \citet{Soler16}, we developed a custom logger instead for general flexibility and to suit our experimental design. Most notably, the logger was designed to record five analog channels continuously at a sampling rate of 1000 Hz to an SD-card, including timestamps at high temporal accuracy (the order of microseconds) by using GPS time synchronization. The latter allows cross correlation between instruments without the need of instruments to communicate with each other or be connected to a centralized system.

The hardware of the instrument consists of five main components: (1) a microcontroller, Arduino Due; (2) SD-card breakout, used to store the recorded signals; (3) GPS breakout, used for time synchronization and recording of its location; (4) signal amplifiers, to amplify the voltage signals generated by (5) the geophone. Both firmware and design of the hardware are made open-source and can be found at http:/$\!$/github.com/jvoermans/Geophone\_Logger.

Here, we used a triaxis GS-One 10 Hz geophone, which has a spurious frequency $>240$ Hz and sensitivity of 85.5 V/m/s. Based on the datasheet, the geophone response is linear for $f>10$ Hz, and decays exponentially for smaller frequencies. The geophones used in this study were not calibrated, meaning that the recorded voltages cannot be converted to surface elevation with sufficient accuracy. However, phase information of the recorded vibrations are unaffected. The $x$ and $y$-components of the geophones were amplified by a gain of 50, whereas the signal of the $z$-component was split and amplified by gains of 5, 50 and 1000 to increase the range of sensitivity of the logger. We note that the z-component of geophone 1 did not work for unknown reasons.

\subsection{Elastic modulus}
\begin{figure}[ht]
	\centering
	\includegraphics{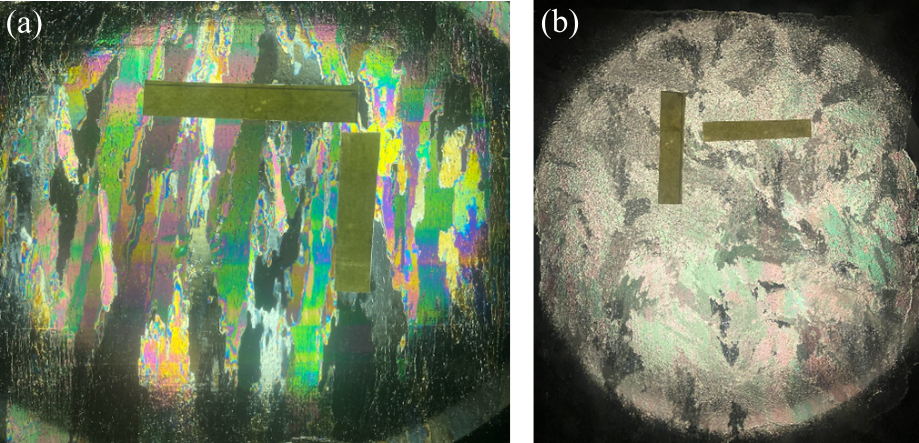}% Images in 100% size
	\caption{Photographs of vertical (a) and horizontal (b) thin sections of sea ice in polarized light. Thin sections were made from a specimen of sea ice from the Templefjorden, March 2022.  Yellow strips scale 5 cm length}
	\label{fig:fig2_ice_columns}
\end{figure}
To provide comparison of our estimates of the effective elastic modulus, in-situ cantilever experiments were performed on the ice to obtain estimates of the effective elastic modulus $E^*$. Eigen frequency of bending oscillations of floating fixed-ends beam depends on the elastic modulus of the beam along the beam axis. The eigen frequency depends also on the added mass of the beam determined by the beam geometry. The fixed-ends beam was cut from sea ice near the ice edge of the Templefjorden on March 9, 2022 (see Fig. \ref{fig:fig1_experiment_setup}). The beam length was $2L_b=3$ m, the beam width was $2b=0.27$ m, and the ice thickness was $H=0.3$ m. The ice temperature was measured between -1.9$^\circ$C and -2.5$^\circ$C, the mean ice salinity was 5 ppt. The ice had columnar structure with the diameter of columns up to 2 cm. Boundaries of ice columns had complicated geometry (Fig. \ref{fig:fig2_ice_columns}). The length of some columns exceeded 7 cm. The water temperature below sea ice was measured as -1.7 C.

\begin{figure}[ht]
	\centering
	\includegraphics{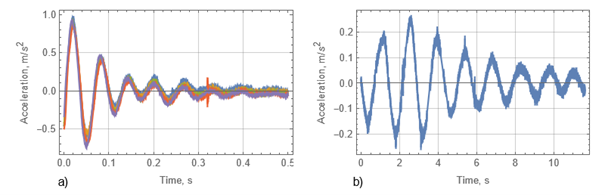}% Images in 100% size
	\caption{Natural oscillations of the beam with fixed ends (a), and the free-floating beam (b).}
	\label{fig:fig3_E_vibrations}
\end{figure}

The accelerometer Bruel \& Kjær DeltaTron Type 8344 designed for the measurements of vibrations in the frequency range 0.2 Hz – 3 kHz was used to measure the eigen frequency of the fixed-ends beam in the vertical direction. It was placed in the middle of the beam surface. The oscillations were initiated by jumping on the ice near the beam. An example of the records after five tests is shown in Fig. \ref{fig:fig3_E_vibrations}(a). One can see that period of oscillations are very similar. The eigen frequency of the beam was found as $\omega=104$ rad/s. After the test, the beam was cut near the ends and the frequency of free heave oscillations of the beam was measured with the same accelerometer. The eigen frequency of the free-floating beam was measured at $\omega_b=4$ rad/s (Fig. \ref{fig:fig3_E_vibrations}b). 

The elastic modulus of a fixed-ends beam is determined by:
\begin{equation}
\begin{array}{l}
E^*=\frac{12(\omega^2 - \omega_b^2)(\rho_i H+m_{add})}{H^3}\frac{L_b}{2.365}\\
\omega_b^2=\frac{\rho_w g}{\rho_i H+m_{add}}
\end{array}
\label{eq:Elastic_modulus}
\end{equation}
where $\omega_b$ is the frequency of heave oscillations of the free-floating beam, $m_{add}$ is the added mass per unit surface of the beam, $g$ is the gravitational acceleration, and $\rho_i$ and $\rho_w$ are the densities of the ice and water, respectively. Based on the observations, the effective elastic modulus was estimated at $E^*=488$ MPa.

\subsection{Data Analysis}
\label{sec:DataAnalysis}
Source triangulation was used to determine where the vibrations originate from and at what speed the vibrations were propagating. The arrival times of vibration events at the location of the geophones may be determined by:
\begin{equation}
\begin{array}{l}
t_1=t_0+\sqrt{(x_1-x_0)^2+(y_1-y_0)^2}/c\\
t_2=t_0+\sqrt{(x_2-x_0)^2+(y_2-y_0)^2}/c\\
t_3=t_0+\sqrt{(x_3-x_0)^2+(y_3-y_0)^2}/c
\end{array}
\label{eq:set_equations}
\end{equation}
where $t$ refers to time and $(x, y)$ the coordinates in a relative frame of reference and $c$ is the propagation speed of the waves. Subscripts refer to the geophone (1--3) or the source (0), i.e., $t_1$ is the instant at which a vibration event arrives at the location of geophone 1 at $(x_1, y_1)$. Eq. \ref{eq:set_equations} contain four unknowns and thus would normally require four geophones to identify where the vibrations are coming from. However, if information is known about the speed at which these vibrations propagate, one more equation is available and the source of the vibration event can be determined with three geophones. In the case of compressive and shear waves, the wave speed is given, respectively, by \citep{Stein98}:
\begin{equation}
\begin{array}{l}
c_c=\sqrt{\frac{E}{\rho_i (1-\theta^2)}}\\
c_s=\sqrt{\frac{E}{2\rho_i (1+\theta)}}
\end{array}
\label{eq:Compression_wave}
\end{equation}
where $\theta$ is the Poisson ratio, here taken as 0.3. For flexural waves in sea ice we assume the ice to be a thin elastic plate such that the dispersion relation of flexural waves in ice can be modeled as \citep{Fox91,Collins17}:
\begin{equation}
\frac{\omega^2}{Lk^4/\rho_w - M\omega^2 +g}=k \tanh(kd) 
\label{eq:dispersion_relation}
\end{equation}
with $L=E^*H^3/(12[1-\theta^2])$, $M=H\rho_i/\rho_w$, where $\omega=2\pi/T$ is the radian frequency with wave period $T$, the wave number $k$, and water depth $d$. From Eq. \ref{eq:dispersion_relation} the propagation speed of wave energy of flexural waves (the group velocity $c_g$) can be determined as $c_g=\partial \omega/\partial k$. We note that Eq. \ref{eq:dispersion_relation} introduces two additional unknowns, the ice thickness $H$ and effective elastic modulus $E^*$. However, as flexural waves are dispersive, observations of $t_1$, $t_2$ and $t_3$ across a wide range of wave frequencies may provide enough information to determine $x_0$, $y_0$, $H$ and $E^*$.

We use here two different approaches in retrieving information of the ice properties and the source of vibration events. We focus for this on flexural waves. The first approach considers the difference in arrival times of vibrations with frequency $f$ between pairs of geophones to estimate the group velocity as $c_{g,ij}(f)=(\sqrt{(x_i-x_0)^2+(y_i-y_0)^2}-\sqrt{(x_j-x_0)^2+(y_j-y_0)^2})/(t_i-t_j)$, where $i$ and $j$ refer to the geophone numbers. For this, arbitrary $(x_0, y_0)$ are taken to estimate $c_g(f)$ for each pair, where solutions for $c_g(f)$ and $(x_0,y_0)$ are found when $c_{g,12}=c_{g,13}=c_{g,23}$. For different $f$, this provides observations of $c_g$ across a range of $f$ to which the dispersion relationship Eq. \ref{eq:dispersion_relation} can be fitted. For the second approach, Eqs. \ref{eq:set_equations} and \ref{eq:dispersion_relation} are solved iteratively to obtain the lowest root-mean-square error (RMSE) in $t_1$, $t_2$ and $t_3$ with variable $(x_0,y_0,t_0)$, $H$ and $E^*$. The first approach may be seen as local as the phase speed of the vibrations are estimated locally at the site of the geophones, whilst the second approach may be viewed as non-local as the sea ice properties between the vibration source and the geophones will influence the results.

\section{Results}
\begin{figure}[ht]
	\centering
	\includegraphics{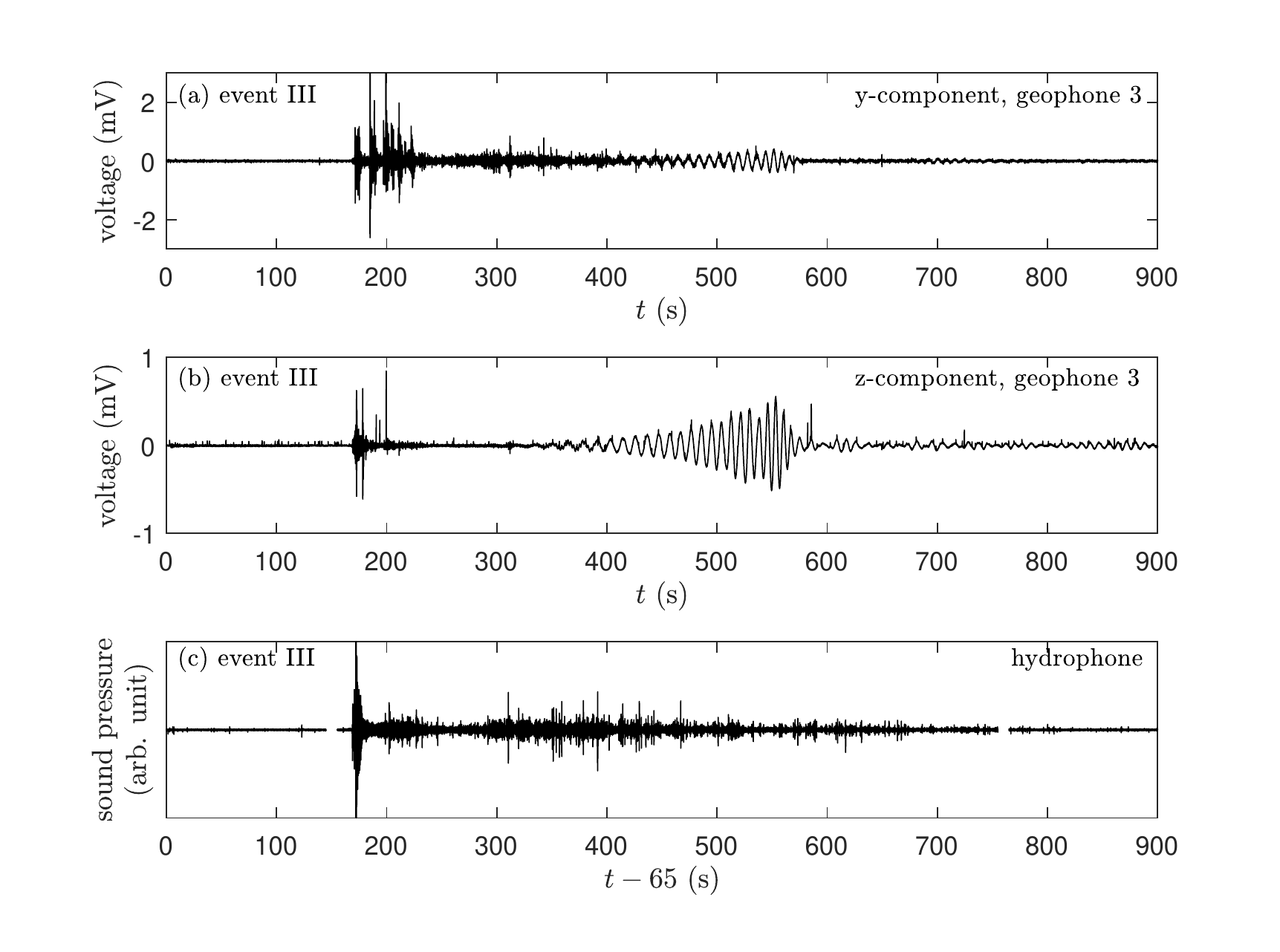}% Images in 100% size
	\caption{Timeseries of (a) horizontal and (b) vertical vibrations recorded by geophone 3 during event III. A high frequency event around $t=200$ s is closely followed by low frequency waves between $t=300$ s and 600 s. Note that the vertical axes are scaled differently. Hydrophone recording during is shown in (c) shifted by 65 s, a shift originating from clock drift of the hydrophone logger, to match the time of the initial vibrations recorded by the geophone loggers.}
	\label{fig:fig4_timeseries}
\end{figure}

In this study we focus on three major events that were recorded on the 18th, 21st and 22nd of February by all three geophones. We refer to these events as event I--III, respectively. In Fig. \ref{fig:fig4_timeseries}a,b the timeseries of the $y$ and $z$ components of geophone 3 during event III are shown. We note that events I and II have similar characteristics as event III, except that the magnitude of the vibrations recorded during event III were the largest. At around $t=170$ s, high frequency vibrations are observed, followed by low frequency waves over the next few hundred seconds. The high frequency vibrations are strongest in the horizontal direction and relatively weak in the vertical component, suggesting this may be a compressive wave. The low frequency waves induces surface velocities similar in magnitude in both horizontal and vertical direction. While magnitude of the voltage amplitudes from $t=400$ to 550 s as observed in Fig. \ref{fig:fig4_timeseries}b is increasing, we would like to reiterate that this is not directly translatable to an increased amplitude of the surface elevation of the ice as the geophones response to low frequencies is not linear. In Fig. \ref{fig:fig4_timeseries}c the corresponding timeseries of sound pressure recorded by the hydrophone is shown. We note that we have shifted the time series of the hydrophone recordings by 65 seconds to match the records of the geophone loggers. This shift is caused by clock drift as the hydrophone logger, unlike the geophone loggers, is not synchronized by GPS. Based on the audio records, we suspect the event to originate from the glacier and caused by calving (a snapshot of the audio record during this initial event can be found at http:/$\!$/doi.org/10.5281/zenodo.7750699). 

\begin{figure}[ht]
	\centering
	\includegraphics{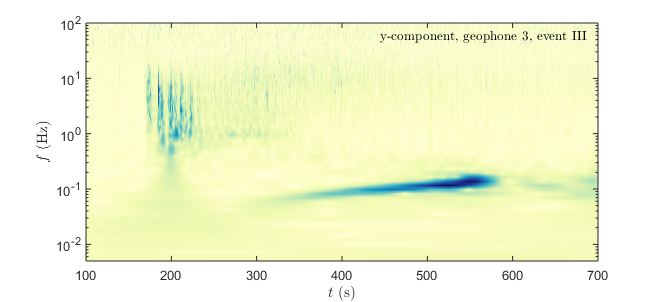}% Images in 100% size
	\caption{Continuous wavelet transform of the timeseries shown in Fig. \ref{fig:fig4_timeseries}a. Colour is indicative of the energy, with blue indicating high energy and yellow low energy.}
	\label{fig:fig5_wavelet}
\end{figure}

\begin{figure}[ht]
	\centering
	\includegraphics{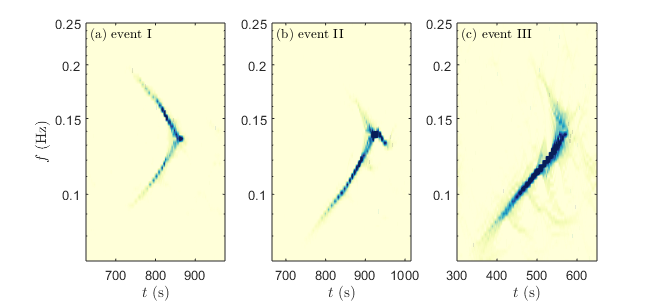}% Images in 100% size
	\caption{Wavelet synchrosqueezed transform of the y-component of geophone 3 during events (a) I, (b) II and (c) III, respectively. Colour is indicative of the energy, with blue indicating high energy and yellow low energy.}
	\label{fig:fig6_wavelet_all}
\end{figure}

In Fig. \ref{fig:fig5_wavelet} the continuous wavelet transform is shown of the timeseries presented in Fig. \ref{fig:fig4_timeseries}a. The initial vibrations around $t=200$ s are within a frequency range of 0.4--18 Hz and do not appear to be dispersive. To obtain information on the times at which these high frequency vibrations arrived at the geophones, we cross-correlate between the horizontal components of geophone pairs to estimate the transit time of the initial high frequency vibrations between these pairs, i.e., this will provide estimates of $t_1-t_2$, $t_1-t_3$ and $t_2-t_3$. Following the first approach (see Data Analysis section), we estimate the propagation speed of these initial vibrations to be $2178\pm40$, $2184\pm20$ and $2321\pm20$ m/s for events I--III, respectively, and are coming from the direction of the glacier. We suspect that these vibrations are compressive waves and thus do not satisfy the frequency dispersion relation of Eq. \ref{eq:dispersion_relation}.

\begin{figure}[ht]
	\centering
	\includegraphics{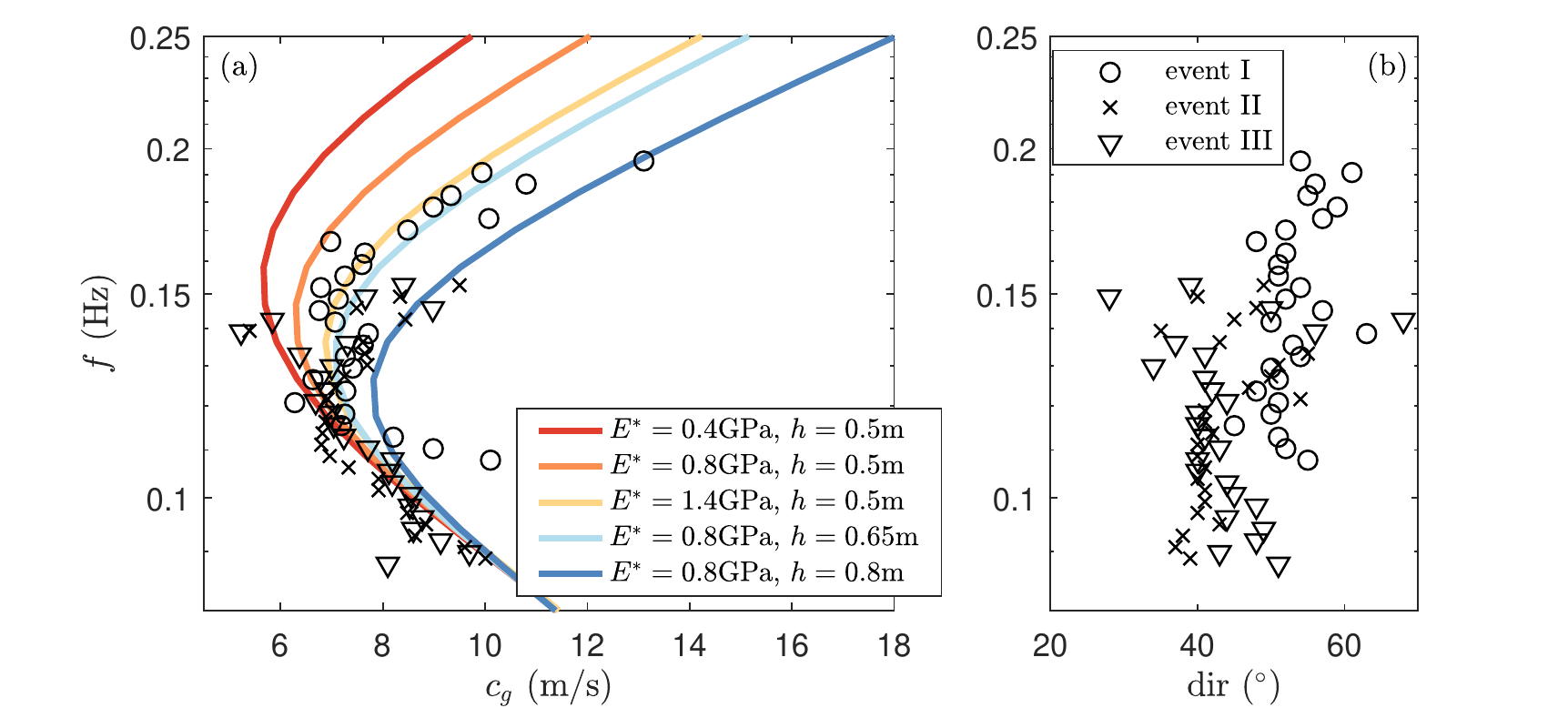}% Images in 100% size
	\caption{(a) Estimates of group velocity as determined by the transit time between the three geophones for the three wave events. In colour are given various estimates of the group velocity based on different effective elastic modulus $E^*$ and sea ice thickness $H$. (b) Estimates of the direction of the wave source relative to the geophones, where North is taken as 0$^\circ$.}
	\label{fig:fig7_Cg}
\end{figure}

As the low frequency waves that follow the initial high frequency vibrations appear to be dispersive (see Fig. \ref{fig:fig5_wavelet}), and the waves are unlikely to have been generated in the open water, we can obtain the $f$-$t$-relationship from the wavelet spectrum. For this, we use the wavelet synchrosqueezed transform using Matlab's \textit{wsst}-function and bump wavelet to identify the times of maximum energy at each frequency (Fig. \ref{fig:fig6_wavelet_all}). We note that minima in $c_g$ (or maxima in $t$) can be observed around 0.14 Hz. From the transit times between each geophone pair and for each wave frequency we can then estimate $c_g$ as a function of $f$. In Fig. \ref{fig:fig7_Cg}a estimates of $c_g$ are compared against the wave dispersion model in sea ice (Eq. \ref{eq:dispersion_relation}) for various values of the ice thickness $H$ and effective elastic modulus $E^*$. For waves with $f\lesssim 0.1$ Hz the propagation speed of waves is largely unaffected by the ice. Notable deviations start to occur for  $f\gtrsim 0.11$ where thicker ice and a larger effective elastic modulus lead to increases in $c_g$. Knowing that the ice thickness near the geophones was measured at about 0.5 m, our initial estimate of $E^*$ is 1.4 GPa. This is, however, strongly based on the observations during event I in the frequency range 0.16--0.2 Hz. From the transit times between geophone pairs, i.e., $t_1-t_2$, $t_1-t_3$ and $t_2-t_3$, we can also identify the direction of the source of the waves relative to the geophones (Fig. \ref{fig:fig7_Cg}b). Assuming that all waves are supposed to come from the same source for each event, and thus same direction for each event, it suggests that uncertainty in estimates of $c_g$ is largest for the higher frequencies.

\begin{figure}[ht]
	\centering
	\includegraphics{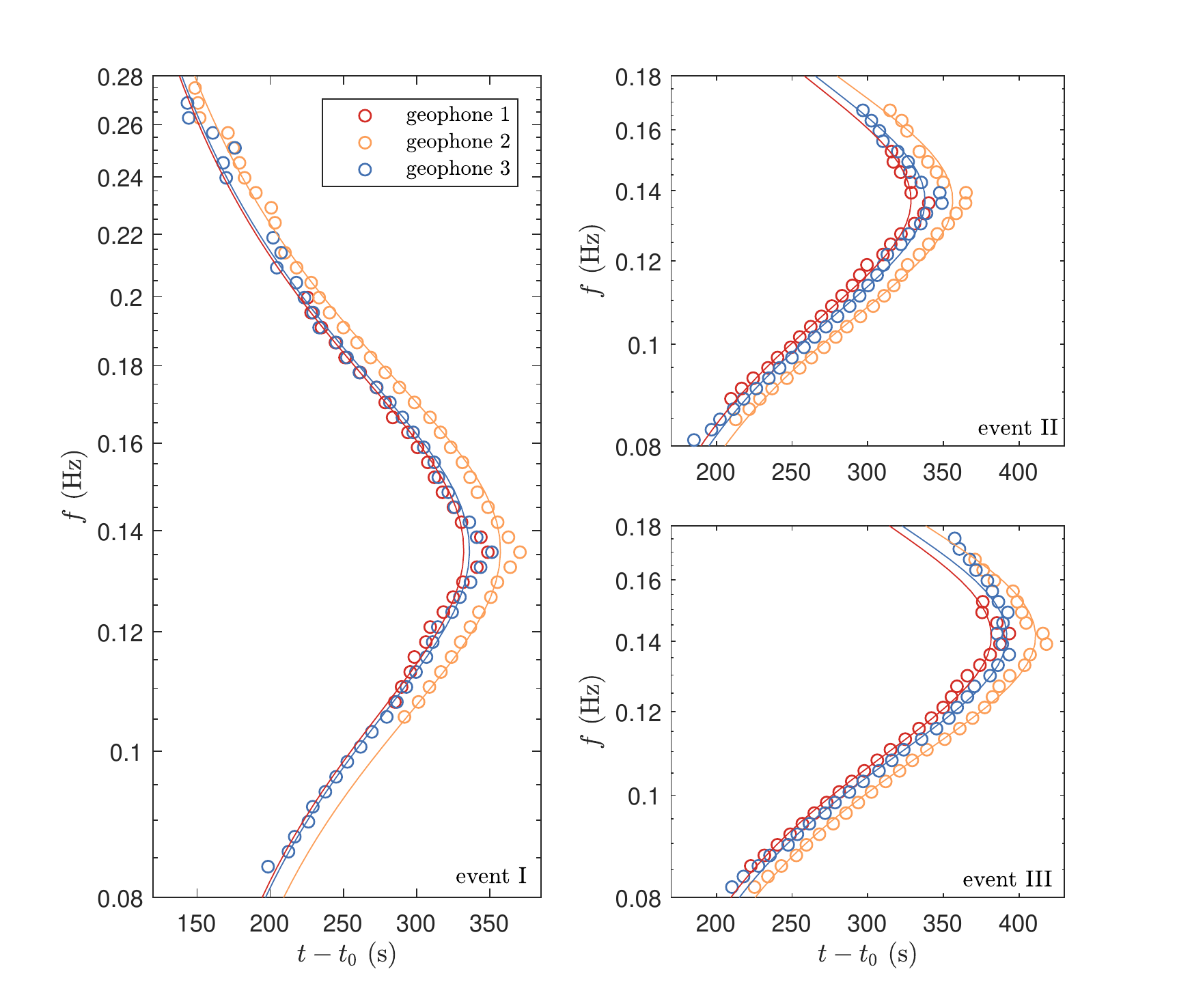}% Images in 100% size
	\caption{Best fits of the dispersion relationship in sea ice (solid line) against observed arrival times at the three geophones for the three events (markers). Note, multiple solutions exist for $H$-$E^*$ that can replicate the best fit, see Fig. \ref{fig:fig9_Y_H}.}
	\label{fig:fig8_profiles}
\end{figure}

In Fig. \ref{fig:fig8_profiles} the dispersion relationship is fitted directly to the observed $f$-$t$-curves following the second approach (see Data Analysis section) to obtain best estimates of $H$ and $E^*$. The observations match the dispersion relationship of waves in ice well, although the minima in $c_g(f)$ (or maxima in $t(f)$) tend to be overpredicted (underpredicted). This may, however, also be an artifact of the wavelet synchrosqueezed transform.

\begin{figure}[ht]
	\centering
	\includegraphics{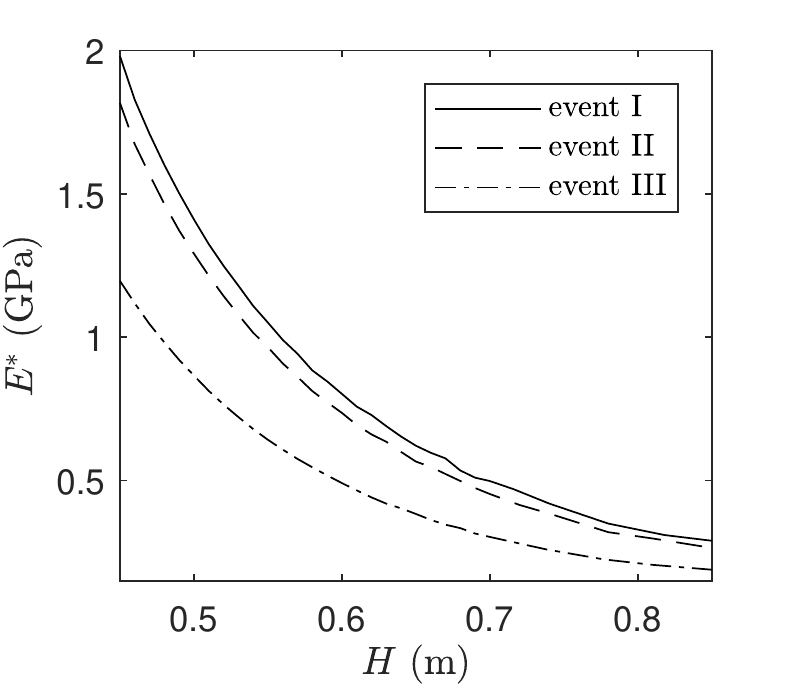}% Images in 100% size
	\caption{Best fit solutions to the wave arrival times as measured by the three geophones for the effective elastic modulus $E^*$ and sea ice thickness $H$.}
	\label{fig:fig9_Y_H}
\end{figure}

As impacts of $E^*$ and $H$ on $c_g$ are similar for frequencies smaller than about $f\lesssim 1$ Hz, multiple combinations of $E^*$ and $H$ can approximate the observations well (Fig. \ref{fig:fig9_Y_H}). Noting that the ice thickness around the geophones was measured at $H\approx 0.5$ m thick, and ice thickness near the glacier may be estimated at about $H\approx 0.75$ m based on extrapolation of the ice thickness measurements (a value similar to those measured near the glacier in 2011, e.g., \citet{Marchenko11}), the average ice thickness through which the vibrations traveled to reach the geophones is roughly 0.63 m. This would result in best estimates of the elastic modulus as 0.67, 0.64 and 0.42 GPa for events I--III, respectively. We note that this is consistent with the measured values of $E\approx 0.5$ GPa at the ice edge from the cantilever experiments (see section Elastic Modulus).

\begin{figure}[ht]
	\centering
	\includegraphics{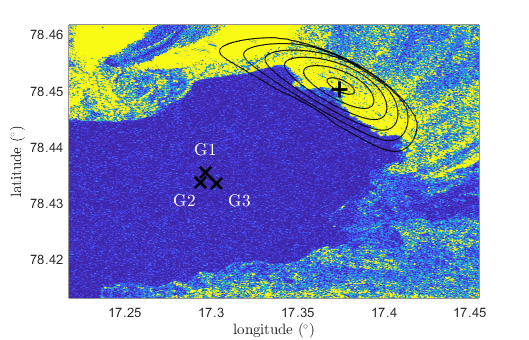}% Images in 100% size
	\caption{Estimates of the source of the dispersive waves during event III (plus sign) and associated contours of the RMSE of the best fits. Location of the geophones are identified by cross-markers. Sentinel-1 image taken on 24-02-2022 is given in colour.}
	\label{fig:fig10_map_contour}
\end{figure}

\begin{figure}[ht]
	\centering
	\includegraphics{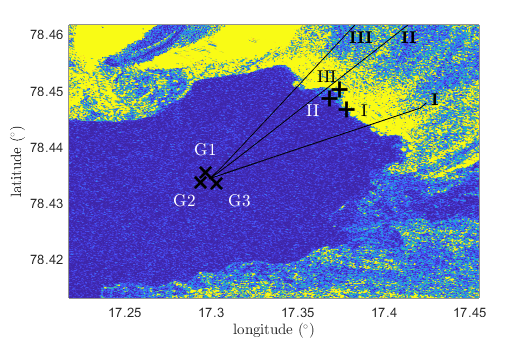}% Images in 100% size
	\caption{Estimates of the source of the dispersive waves during events I--III (plus markers) and direction of the high frequency events (solid lines). Location of the geophones are identified by cross-markers. Sentinel-1 image taken on 24-02-2022 is given in colour.}
	\label{fig:fig11_map}
\end{figure}

Aside from estimates of the ice thickness and effective elastic modulus, the source of the vibrations can be approximated by finding the lowest RMSE fit of the dispersion relation against the $f$-$t$-curve. Specifically, we find a global minimum for $(x_0,y_0)$ for all three dispersive wave events. Contours of the RMSE for event III are shown in Fig. \ref{fig:fig10_map_contour} with a best estimate location of the vibration source located near the glacier wall, confirming that the events are originating from the glacier. In Fig. \ref{fig:fig11_map} the best-estimate source locations of all three events are shown, including the direction of the high-frequency vibrations. We note that the direction of the high frequency vibration is based on a single data point and is thus expected to be less accurate than the estimate based on the low-frequency waves. They are, nevertheless, consistently directed towards the glacier.

\section{Discussion}
In this study we estimated the effective elastic modulus and ice thickness from passive wave observations. We find that the effective elastic modulus derived from fitting the observed wave arrival times to the modeled arrival times retrieves estimates of 0.4--0.7 GPa, whereas estimates based on observations of the group velocity leads to a value of approximately 1.4 GPa. We believe the former is more accurate given that it uses considerably more data to perform the fitting. Particularly, we see scatter in the direction estimates of the vibration sources relative to the instruments for the higher frequency range (i.e., Fig. \ref{fig:fig7_Cg}b) which is a frequency range fundamental in the fit against the dispersion relationship in ice. Moreover, the measured effective elastic modulus of 0.5 GPa from cantilever experiments is consistent with the former as well. Nevertheless, the difference may also be attributed to the variability of sea ice conditions, in particular the sea ice temperature. The sea ice temperature near the ice edge was measured between -2.5$^\circ$C and -1.5$^\circ$C, and expected to be much lower near the glacier between -15$^\circ$C and -10$^\circ$C. Considering either of these estimates, our results suggests that the usage of a default value of $E^*=5.5$ GPa in wave-ice interaction models may lead to significant errors and/or uncertainties in model simulations as those measured in this study are significantly lower. Further field experiments need to be performed to provide more information on the  variability of $E^*$ in the field and to provide further recommendations on how to parameterize $E^*$.

\begin{figure}[ht]
	\centering
	\includegraphics{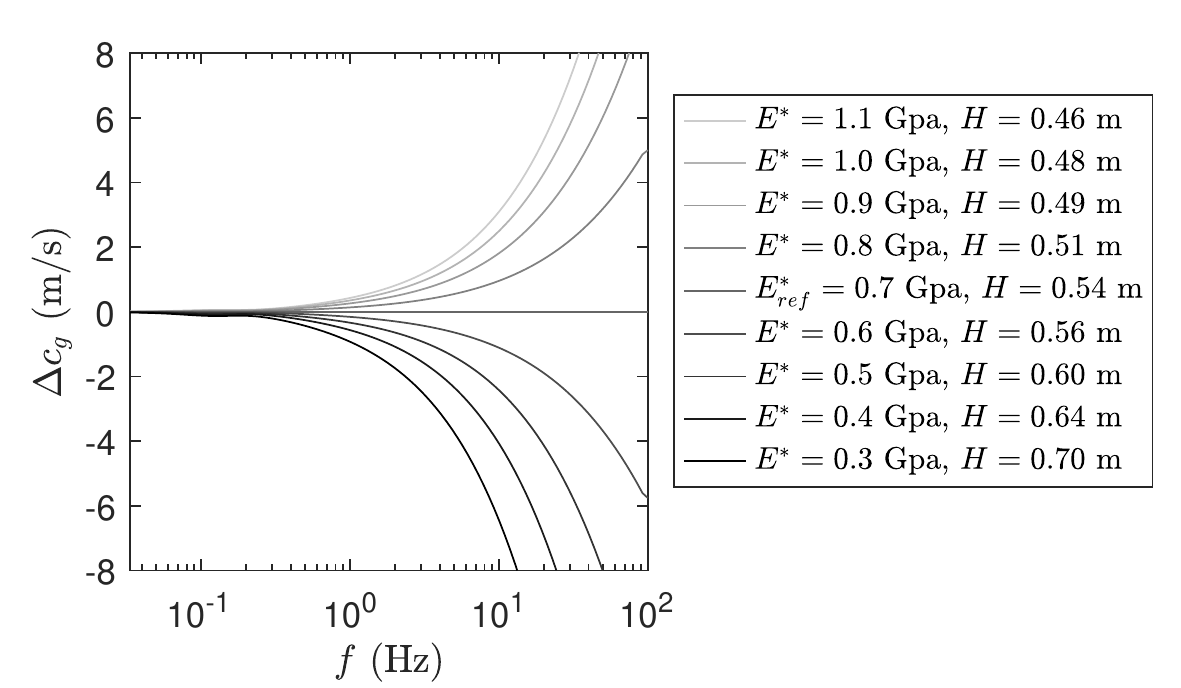}% Images in 100% size
	\caption{Variability of the group velocity $\Delta c_g=c_g-c_{g,ref}$ for combination solutions of $H$-$E^*$ with wave frequency. Here, $E^*=0.7$ GPa and $H=0.54$ m are taken as for the reference velocity $c_{g,ref}$. Impact of $E^*$ and $H$ to the shape of the dispersion relationship become significant for frequencies above 1 Hz.}
	\label{fig:fig12_10Hz}
\end{figure}

\begin{figure}[ht]
	\centering
	\includegraphics{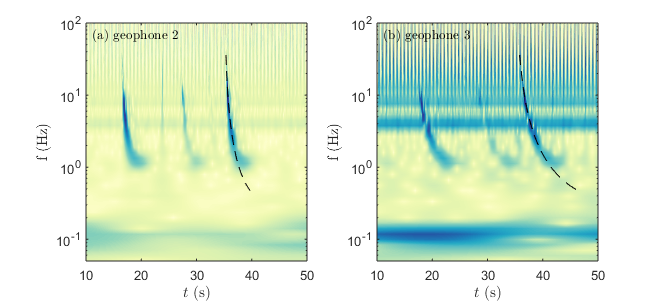}% Images in 100% size
	\caption{Example wavelet spectrum of the z-component of the geophone loggers during suspected cracking events. Black dashed line indicates best fit. Colour is indicative of the energy, with blue indicating high energy and yellow low energy.}
	\label{fig:fig13_wavelet_crack}
\end{figure}

While our observations were able to provide estimates of $H$-$E^*$ combinations, details on the ice thickness were required before $E^*$ could be estimated explicitly. This is largely because the impacts of $E^*$ and $H$ on the shape of the dispersion relationship are too small for the frequency range of our observations from 0.08--0.28 Hz. To illustrate, in Fig. \ref{fig:fig12_10Hz} we compare $\Delta c_g=c_g-c_{g,ref}$ for different points on the $H$-$E^*$ curve for event III (see Fig. \ref{fig:fig9_Y_H}), with $E^*=0.7$ GPa and $H=0.54$ m taken as the reference group velocity $c_{g,ref}$. Up to $f\approx 1$ Hz the different combinations of $H$-$E^*$ have very limited impact on the shape of $c_g(f)$, meaning that the absolute differences in $c_g$ for different $H$-$E^*$ combinations are well within the measurement uncertainty of $c_g$. However, if observations are available for $f>1$ Hz, the range of $H$-$E^*$ solutions may be reduced significantly as differences in the group velocities between different solutions may exceed 1 m/s, whilst for observations of $f>10$ Hz we expect that explicit estimates of $E^*$ and $H$ become feasible. Although the dispersive waves presented in this study (i.e., Fig. \ref{fig:fig4_timeseries}) does not contain significant or identifiable energy at $f>0.3$ Hz, we note that measurements of dispersive waves at a higher frequency range are frequently encountered in our dataset. For instance, in Fig. \ref{fig:fig13_wavelet_crack} an example of vibrations caused by a sequence of suspected (thermal) cracks is shown with identifiable phase-time information extending beyond 10 Hz. Unfortunately, no further information of $H$-$E^*$ could be obtained from this data here as these cracking signals were only identifiable in the z-components of the geophones whilst only two geophones had this component working. Nevertheless, from the continuous wavelet transform we can see that the events arrived first at geophone 2 and we can infer from the data that, taking $E^*=0.6$ GPa and $H=0.52$ m, the source of these cracking events was about 110 m from geophone 2 and 260 m from geophone 3 (see dashed lines in Fig. \ref{fig:fig13_wavelet_crack}).

\begin{figure}[ht]
	\centering
	\includegraphics{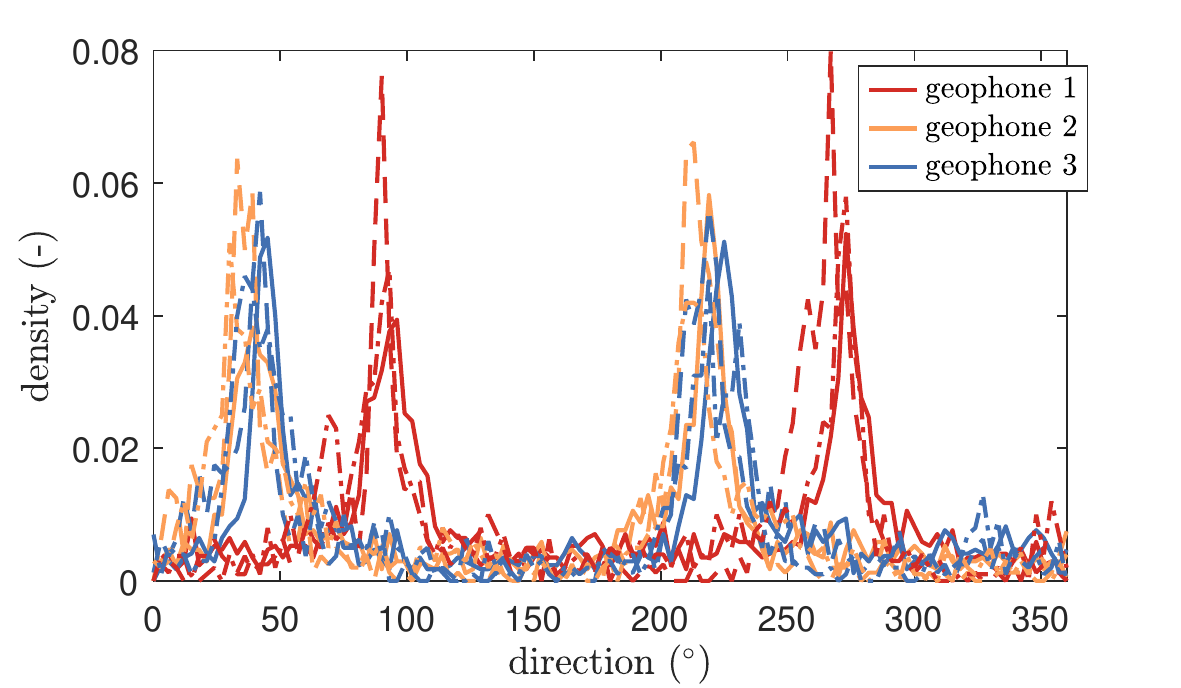}% Images in 100% size
	\caption{Probability density of the vector direction of initial high frequency vibrations during event I (solid line), event II (dashed line) and event 3 (dash-dot line). North and east correspond to 0$^\circ$ and 90$^\circ$, respectively.}
	\label{fig:fig14_Vector}
\end{figure}

Observations of the elastic modulus thus far referred to the effective elastic modulus. However, from the propagation speed of compressive and shear waves in the ice, one may estimate the purely elastic modulus as well (i.e., Eq. \ref{eq:Compression_wave}). From the direction of the $x$-$y$ velocity vectors of the geophone signals we may approximate the direction of the initial high frequency vibration events (Fig. \ref{fig:fig14_Vector}). We note that the directions as observed by geophone 2 and 3 are well aligned with the direction from the geophones to the glacier for all three events which supports the hypothesis that the recorded high frequency vibrations are indeed compressive waves. However, the direction of the vectors for geophone 1 during these events is consistently shifted by 50$^\circ$--60$^\circ$ for unknown reasons, but we suspect that this is because an error in its alignment during deployment. From Eq. \ref{eq:Compression_wave} the purely elastic modulus $E$ is then estimated at $3.98\pm0.14$, $4.00\pm0.07$ and $4.52\pm0.08$ GPa for events I--III, respectively. We note that the purely elastic modulus is always larger than the effective elastic modulus, except for very high loading rates where $E^*$ will approach the value of $E$. We thus find here that $E\approx 6 E^*$ under the conditions experienced during the field experiment, which is similar to the results of \citet{Karulina19} who observed a difference by a factor of 4-5 between measurements of the elastic modulus obtained through cantilever experiments and direct acoustic observations.

Based on the geophone and hydrophone records, we suspect events I--III to originate from calving events at the glacier wall. The number of reported ice failure events in Spitsbergen glaciers is estimated to be in the range of 50-100 in February based on observations in the years 2010-2012 \citep{Fedorov16}, with most events not related to glacier wall calving and occur in the accumulation zone of glaciers. This seems to be consistent with our observations where only three significant events were recorded over a period of about two weeks. From this, we may also conclude that ice failure events at a distance from glacier front do not bring energy to the sea ice near the wall.

Although our experiments provide estimates of the effective elastic modulus similar to that measured in-situ through cantilever experiments, there are limitations to the methodology used here. Most importantly, we assume here that the wave dispersion model of \citet{Fox91}, i.e., Eq. \ref{eq:dispersion_relation}, is valid. Secondly, our observations are most likely tsunami events caused by glacier calving, in analogy to landslide induced tsunamis. The generation of such long waves at such proximity of the instruments are restricted to very limited sites across the polar regions. When waves are generated in open water and cross various types of sea ice, the cumulative impact of the inhomogeneous ice on the group velocity of low frequency waves makes it problematic to estimate the effective elastic modulus and ice thickness based on the arrival times of waves (i.e., second approach, see Data Analysis section). In such a case, $E^*$ can only be measured by estimating $c_g$ from the transit times between the instruments (i.e., first approach, see Data Analysis section) or from high frequency vibrations. Noting that a 1 s period wave in sea ice has a wave length of roughly 20 m, the main length scale restriction for an ice floe then seems to be the minimum distance required between the geophones to measure the travel times of vibrations between the geophones. Based on our observations here we expect this minimum distance to be about 100 m and thus a minimum diameter of an ice floe of a few hundred meters. While we used three geophones in this study, we recommend to increase this to 4--5 geophones in future experiments, which is expected to increase the accuracy of estimates of sea ice properties and safeguards project outcomes in case of instrument malfunctioning. Further development of the geophone logger is required for automatic detection of significant vibration events and the corresponding dispersion curves when instruments cannot be retrieved. This would require developments in on-board data processing capabilities and changes in hardware to accommodate for this (such as the addition of a microprocessor like the Raspberry Pi). While the most promising route for automatic signal detection seems to be through wavelet spectrum analysis (e.g., see Figs. \ref{fig:fig6_wavelet_all} and \ref{fig:fig13_wavelet_crack}), further study is required in establishing threshold criteria for the fitting of $f$-$t$-curves and measures of quality control when there is no access to the raw data.

\section{Conclusions}
We showed that the elastic modulus and sea ice thickness of fast ice can be estimated from measurements of waves in sea ice. By determining the arrival time of wave events at the geophones and their phase speed, we obtain estimates of the purely elastic modulus during three different major wave events of 4--4.5 GPa, and the effective elastic modulus of 0.4-0.7 GPa. Estimates of the effective elastic modulus correspond well to the measured effective elastic modulus of 0.5 GPa from in-situ cantilever experiments. Unfortunately, no explicit estimates of the elastic modulus and ice thickness were possible with our current dataset as their respective impacts on the shape of the dispersion relationship is similar for frequencies below about 1 Hz. Thus, in-situ measurements of ice-thickness were required to obtain explicit estimates of the effective elastic modulus. However, explicit estimates of both ice thickness and elastic modulus will be possible with observations of dispersive waves with frequencies larger than 1 Hz. We note that such signals are frequently observed in our dataset, but due to experimental limitations could not be used here. Nevertheless, our results confirm that low-cost geophones can be used to estimate sea ice properties in fast ice passively. Based on the frequent occurrence of high frequency dispersive wave events in our dataset beyond 10 Hz, we expect this method to be possible on large ice floes as well. Progress in retrieving sea ice properties from wave measurements is expected to increase the dataset of in-situ sea ice properties observations and improve our understanding of wave-ice interactions.

\section*{Acknowledgements}

Alexander V. Babanin acknowledge support from the US Office of Naval Research grant N62909-20-1-2080. Joey J. Voermans, Alexander V. Babanin were supported by the Australian Antarctic Program under project 4593. Joey J. Voermans, Jean Rabault, Aleksey Marchenko, Takuji Waseda, and Alexander V. Babanin acknowledge the support of the Research Council of Norway and Svalbard Science Forum under project 311266. Authors would like to acknowledge that this work contains modified Copernicus Sentinel data 2022 processed by Sentinel Hub. The acoustic recording of the hydrophone during event III can be found at http:/$\!$/doi.org/10.5281/zenodo.7750699. Firmware and hardware design is made open-source and can be found at http:/$\!$/github.com/jvoermans/Geophone\_Logger. All data is made available in an online repository, please visit the github repository for further directions on accessing the data.

%\clearpage

\bibliography{templateArxiv}

\end{document}